\documentclass[12pt,a4paper,aps,preprint,superscriptaddress,nofootinbib]{revtex4-1}
\usepackage[utf8]{inputenc}
\usepackage{graphicx}
\usepackage{amssymb}
\usepackage{textcomp}
\usepackage{amsmath}
\usepackage{tabularx}
\usepackage{bm}
\usepackage{times}
\usepackage{color}
\usepackage{ulem}
\usepackage{slashed}
\usepackage{multirow}
\usepackage{verbatim}
\usepackage{cancel}
\usepackage{subfigure}
\usepackage{epstopdf}
\usepackage{mathrsfs}
\usepackage{amsmath}

\usepackage[colorlinks=true, pdfstartview=FitV, linkcolor=blue, citecolor=blue, urlcolor=blue]{hyperref}
\allowdisplaybreaks[4]

\def\lsim{\mathrel{\raise.3ex\hbox{$<$\kern-.75em\lower1ex\hbox{$\sim$}}}}
\def\gsim{\mathrel{\raise.3ex\hbox{$>$\kern-.75em\lower1ex\hbox{$\sim$}}}}

\definecolor{orange}{rgb}{1,0.5,0}

\begin{document}

%\title{$N_{\rm eff}$  and  the Pseudo-Dirac DM in Higgs portal}
\title{$\mathbf{{N}_{eff}}$  from Excited DM state}

\author{Wei Chao}
\email{chaowei@bnu.edu.cn}
\affiliation{
Center for Advanced Quantum Studies, Department of Physics,
Beijing Normal University, Beijing 100875, China
}
\author{Jing-Jing Feng}
\email{fengjj@mail.bnu.edu.cn}
\affiliation{
Center for Advanced Quantum Studies, Department of Physics,
Beijing Normal University, Beijing 100875, China
}
\author{Ming-Jie Jin}
\email{jinmj@bnu.edu.cn}
\affiliation{
Center for Advanced Quantum Studies, Department of Physics,
Beijing Normal University, Beijing 100875, China
}

\begin{abstract}
For a cold dark matter (DM) originating from the co-annihilation processes, there will be excited state(s) in the dark sector, that may decay or annihilate away shortly after their freeze-out. In this paper we investigate the impact of the decay of these excited states on the effective number of neutrino species, $N_{\rm eff}^{}$, which is an important cosmological parameter and will be tested by the future CMB-S4 project. We work in the framework of pseudo-Dirac DM mode via the Higgs portal. The relic density and the direct detection signal of the DM are calculated, which gives the available parameter space. Impacts of the excited state (the heavy component of the pseudo-Dirac fermion) on the decoupling of active neutrinos are investigated by solving Boltzmann equations of photon and neutrinos with collision term induced by the decay of the excited state. The numerical result shows that $|N_{\rm eff}^{}|$ can be of the order $10^{-4}$, which depends on the mass and lifetime of excited DM state.
\end{abstract}

\maketitle

%%%%%%%%%%%%%%%%%%%%%%%%%%%%%%%%
\section{Introduction}
\label{sec:Intro}
%%%%%%%%%%%%%%%%%%%%%%%%%%%%%%%%

Various cosmological observations have confirmed the existence of dark matter (DM)\cite{Albada:1986roa,Efstathiou:1986bf,Springel:2005nw,WMAP:2008lyn}, which should couple to the Standard Model (SM) particles so as to interpret its production and evolution in the early Universe.  However, what is DM made by and how it couple to the SM still elude us, which catalyzed  model buildings of DM. Of various DM candidates, the weakly interacting massive particle (WIMP)~\cite{Jungman:1995df} is one of the most promising DM candidate as it can naturally explain the observed relic density with its mass at the electroweak scale and its coupling the same as weak coupling.   Moreover, the electric, photon or phonon signals induced by the elastic/inelastic WIMP-nucleon scattering can be coherently enhanced and are testable in  underground laboratories\cite{CDMS:2013juh,CoGeNT:2012sne,PandaX-4T:2021bab}.

A WIMP, namely $\chi$, is in thermal equilibrium with the thermal bath at the early Universe, and freezes out as its interactions rate with the SM particles drops below the expansion rate of the Universe, $n_\chi \sigma < H$, where $n_\chi$ is the number density of  $\chi$, $\sigma$ is the annihilation cross-section and $H$ is the Hubble rate\cite{Edward}. For real cases, there might be excited DM states in the dark sector, which can be heavy neutralino states in supersymmetry~\cite{Jungman:1995df}, Kaluza-Klein tower states in extra dimension theory~\cite{Cheng:2002ej,Servant:2002aq} or other electroweak components in the minimal DM theory~\cite{Cirelli:2005uq}. The relic abundance of DM may come from co-annihilation processes~\cite{BINETRUY1984285,Griest:1990kh,Edsjo:2003us}, which may lower the coupling strength of DM to the SM particle, addressing the non-observation of any direct detection signal in underground laboratories. Since the excited DM decays away in the early Universe, it is impossible to detect its signal in the intensity frontier. It can be detected in the energy frontier by producing this particle directly. A typical signal for this kind of particle is displaced vertex plus missing energy. However, it suffers from the various backgrounds at the LHC~\cite{Ruderman:2011vv}. What is the possible signal of this particle in the comic frontier and how it influences the evolution of the Universe is still elusive to us.

In this paper we study the impact of the excited DM state on the effective number of neutrino species, which is an important cosmological parameter. We work in the framework of pseudo-Dirac DM via the Higgs portal, where the pseudo-Dirac can be decomposed into two Majorana eigenstates, $\chi_{1,2}^{}$ with nearly degenerate masses. We assume $\chi_1$ is the heavy eigenstate with its lifetime depending on the mass splitting of the two eigenstates. $\chi_1$ is first thermalized in the early universe, then  decays into $\chi_2$ and SM particles shortly after its freeze-out. We calculate the relic density of $\chi_2$ by solving the  Boltzmann equations of $\chi_{1}$ and $\chi_{2}$ simultaneously and study its signal in the direct detection experiments, from which we derive the available parameter space of the model. Then we calculate the effective number of neutrinos by solving Boltzmann equations that govern the temperature evolution of active neutrinos and photon. Our results show that if $\chi_1$ totally decay into diphoton and $\chi_2$, $\Delta N_{\rm eff}^{} \equiv N_{\rm eff}^{} - N_{\rm eff}^{SM}$ will be a negative value of the order ${\cal O} (10^{-4})$ by taking the DM mass to be ${\cal O}(100~{\rm GeV})$. Alternatively, if $\chi_2$ decays into both neutrinos and diphoton, the sign of $\Delta N_{\rm eff}^{}$ depends on branching ratios of two channels. It should be mentioned that $\Delta N_{\rm eff}^{}$ given in this paper is from excited DM state at the electroweak scale. It can be enlarged to a value detectable by the future CMB-S4 experiment~\cite{Wu:2014hta,Abazajian:2019eic} for a relatively large mass splitting and small DM mass.

The remaining of the paper is organized as follows: In section \uppercase\expandafter{\romannumeral2} we describe the pseudo-Dirac model in detail. Section \uppercase\expandafter{\romannumeral3} is devoted to the calculation of DM relic density as well as its direct detection signal. In section \uppercase\expandafter{\romannumeral4} we calculate the effective number of neutrinos induced by the decay of excited DM states. The last part is concluding remarks. Details of the calculation are listed in Appendices.

%%%%%%%%%%%%%%%%%%%%%%%%%%%%%%%%%%%%%%%%%%%%%%
\section{The model}
\label{sec:Model}
We work in the framework of the pseudo-Dirac DM model. The pseudo-Dirac fermion $\chi$ splits into two non-degenerate Majorana fermion mass eigenstates $\chi_1$ (heavy) and $\chi_2$ (light), with masses $m_1$ and $m_2$ respectively, and mass splitting is defined as $\Delta m=m_1-m_2$. $\chi_2$ is the stable particle with no charge, so it can be a candidate for cold DM\cite{Drees:1992am,Feng:2010gw}. Assuming $\chi$ couples to the SM via the Higgs portal, the Lagrangian can be written as
\begin{eqnarray}
-{\cal L} \sim y \overline{\chi}_L   \Phi  \chi_R^{} + \mu^\prime \overline{\chi_L^{} } \chi_L^C + {\rm h.c.}, \label{L}
\end{eqnarray}
where $\Phi$ is an electroweak scalar singlet coupling to the SM Higgs, $y$ is the Yukawa coupling, $\mu^\prime$ is a tiny Majorana mass.  As $\Phi$ develops a non-zero vacuum expectation value (VEV), $\chi$ gets both Dirac and Majorana mass terms. To derive interactions of the dark sector in the mass eigenstates, we first study the physical parameters of the scalar sector induced by the following Higgs potential,
\begin{eqnarray}
\begin{aligned}
V\left(H,\Phi\right)=&-\mu^{2}H^{\dagger} H+\lambda\left(H^{\dagger} H\right)^2-\mu^2_{\Phi} \Phi^{\dagger} \Phi+\lambda_{\Phi}\left(\Phi^{\dagger} \Phi\right)^2+\lambda_1 \left(\Phi^{\dagger} \Phi\right) \left(H^{\dagger} H\right)\\
&-\mu^{'2}_{\Phi} \Phi^2-\mu^{'2}_{\Phi} (\Phi^{\dagger})^2, \label{eq:vscalar}
\end{aligned}
\end{eqnarray}
where $H=(v_h+ h+i G)/\sqrt{2}$ and $\Phi= (v_\Phi + s+ i a )/\sqrt{2}$ with $v_h$ and $v_\Phi$ the VEVs of $H$ and $\Phi$, respectively.  There are six free parameters left after the symmetry breaking: $m_{\hat h}$, $m_{\hat s}$, $m_{\hat a}$, $v_h$, $v_\Phi$ and $\alpha$, where $m_{\hat{h}}$ is the mass of the SM Higgs $\hat h$, $m_{\hat{s}}$ is the mass of the CP-even scalar $\hat s$, $m_{\hat{a}}$ is the mass of the CP-odd scalar $\hat a$, $\alpha$ is the mixing angle between $\Phi$ and $H$. It should be mentioned that $\alpha $ is strongly constrained by the precision measurements~\cite{ALEPH:2010aa} and the Higgs measurements at the LHC~\cite{ATLAS:2016neq}.

After the symmetry breaking, the mass term for $\chi$ can be written as
\begin{eqnarray}
\frac{1}{2}
\left(
  \begin{array}{cc}
  \overline{\chi_L} & \overline{{\chi_R}^c}
  \end{array}
\right)
\left(
  \begin{array}{cc}
\mu^{'} & m_D\\
m_D     & 0
  \end{array}
\right)
\left(
  \begin{array}{c}
  {\chi_L}^c\\
  \chi_R
  \end{array}
\right)+{\rm h.c.},
\end{eqnarray}
where $m_D= y v_\Phi/\sqrt{2}$ is the Dirac mass.  The mass matrix given in the last equation can be diagonalized by a $2\times 2$ unitary transformation, ${\cal U}^\dagger {\cal M} {
\cal U} ={\rm diag} \{ m_1, m_2\}$ with
\begin{eqnarray}
\cal{U}=\left(
  \begin{array}{cc}
c & -i s\\
s     & i c
  \end{array}
\right),
\end{eqnarray}
where $c=\cos\theta$ and $s=\sin\theta$, with $\theta$ the mixing angle. The phase in the ${\cal U}$ guarantees that mass eigenvalue is positive. To be explicit, one has
\begin{eqnarray}
\begin{aligned}
&m_1=m_D \sin2\theta+\mu^{'}c^2, \\
&m_2=m_D \sin2\theta-\mu^{'}s^2, \\
&\tan2\theta=\frac{2m_D}{\mu^{'}},
\end{aligned}
\end{eqnarray}
which implies that $\theta \approx 45^\circ$ and $\Delta m \approx \mu^\prime$.

Taking $\hat \chi_1$ and $\hat \chi_2$ as the mass eigenstates,  the portal interactions take the following form
\begin{eqnarray}
\begin{aligned}
y \overline{\chi_L} \Phi \chi_R + h.c.
=&\frac{y}{2} \frac{\sin{2\theta}}{\sqrt{2}}\overline{\hat{\chi}_{1}}\left(s^{'} \hat{h}+c^{'} \hat{s}+i\gamma^5\hat{a}\right) \hat{\chi}_{1}+\left(\hat{\chi}_{1}\rightarrow{\hat{\chi}_{2}}\right)\\
&+ y \frac{ \cos{2\theta}}{\sqrt{2}} \overline{\hat{\chi}_{1}}
                                 \left[-i\gamma^5\left(s^{'} \hat{h}+c^{'} \hat{s}\right)+\hat{a}\right] \hat{\chi}_{2}, \label{coupling}
\end{aligned}
\end{eqnarray}
which will be applied to study the phenomena induced by the dark sector. For simplicity, we define $\sin\alpha=s^{'}$ and $\cos\alpha=c^{'}$.

\section{Relic density}
\label{sec:relic}

The $ \chi_1$ and $ \chi_2$ are in equilibrium with the thermal bath in the early Universe and they freeze out when the relevant interaction rates fall below the expansion rate~\cite{Scherrer:1985zt}. Considering that the mass splitting is very small, co-annihilation cannot be ignored when calculating the relic abundance\cite{Gondolo:1997km,Gondolo:1990dk,Profumo:2004wk,Nihei:2002ij}. Quantitively, the  Boltzmann equation for the $i$th spice is
\begin{eqnarray}
\frac{dn_i}{dt}+ 3Hn_i=-\sum^{N}_{j=1}\langle \sigma_{ij}v_{ij}\rangle \left( n_i n_j-n^{eq}_i n^{eq}_j \right),
\label{Boltzmann1}
\end{eqnarray}
where $\langle \sigma_{ij} v_{ij} \rangle $ is the thermal average of the reduced annihilation cross-section of the process $i+j\to {\rm SM} + {\rm SM}$, $n_i^{\rm eq}$ is the equilibrium number density of $i$th particle.
Since all $ \chi_1$ will eventually decay into $ \chi_2$, the final relic abundance can be simply written as  $\Omega_{\rm DM} = m_2 (n_1+n_2)/\rho_c$ with $\rho_c =1.05 \times 10^{-5} (h^2)~{\rm GeV/cm^3}$~\cite{ParticleDataGroup:2020ssz,Bauer:2017qwy} being the critical density.

Defining $n=\sum^2_{i=1}n_i$\cite{Nihei:2002sc,Griest:1990kh},  one can get
\begin{eqnarray}
\frac{dn}{dt}=-3Hn-\langle \sigma_{\rm eff}v \rangle \left(n^2-n^2_{eq}\right)\; ,
\label{Boltzmann}
\end{eqnarray}
where $\langle \sigma_{\rm eff}v \rangle$ can be rewritten into the following convenient form for subsequent calculation and analysis~\cite{Gondolo:1990dk,Srednicki:1988ce},
\begin{eqnarray}
\langle \sigma_{\rm eff}v \rangle=\frac{\int^{\infty}_{4m^2_{DM}} ds s^{3/2}K_1\left(\frac{\sqrt{s}}{T}\right)\sum^N_{ij}\beta^2_f\left(s,m_i,m_j\right)g_i g_j \sigma_{ij}(s)}{8T\left[\sum_i g_i m^2_i K_2\left(\frac{m_i}{T}\right)\right]^2},
\label{sigamav}
\end{eqnarray}
where $s=\left(p_i+p_j\right)^2$ is the usual Mandelstam variable, $K_i$ denotes the modified Bessel function of the order $i$, $g_i$ and $g_j$ are internal degrees of freedom, and $\beta_f(s,m_i,m_j)$ is  the kinematic factor given by
\begin{eqnarray}
\beta_f\left(s,m_i,m_j\right)=\left[1-\frac{\left(m_i+m_j\right)^2}{s}\right]^{1/2}\left[1-\frac{\left(m_i-m_j\right)^2}{s}\right]^{1/2}.
\end{eqnarray}
The total co-annihilation cross section $\sigma_{ij}$ in Eq.(\ref{sigamav}) is calculated as follows
\begin{eqnarray}
\sigma_{ij}=\sum^2_{ij}\sigma\left[\chi_i \chi_j\rightarrow \bar{f}f\left(VV,SS\right)\right],
\end{eqnarray}
where $VV$ represent $ZZ$ and $W^{+}W^{-}$, and $SS$ represent $\hat{h}\hat{h}$, $\hat{s}\hat{s}$, $\hat{a}\hat{a}$. The deduction of $\sigma_{ij}$ is listed in the appendix \ref{appendix A}.

Defining $Y=n/s$, where $s$ is the entropy density, we  can solve the Eq.~(\ref{sigamav}) to get~\cite{Bertone:2004pz}
\begin{eqnarray}
Y_\infty=\left(\sqrt{\frac{\pi}{45}}\sqrt{g_*}m_{\chi}M_{pl}\int^{\infty}_{x_f} \langle \sigma_{eff}v\rangle dx\right)^{-1}\; ,
\end{eqnarray}
where $M_{pl}=1.22\times10^{19}\rm GeV$ is the Planck mass, $x_f=m/T_f$ with $T_f$ being the freeze-out temperature.  Therefore, the relic abundance today in units of the critical density is then given by $\Omega_{DM}= m_2 s_0 Y_\infty /\rho_c $ where $s_0$ is the entropy density today.

\begin{figure}[t]
\begin{center}
\includegraphics[height=6cm,width=8cm]{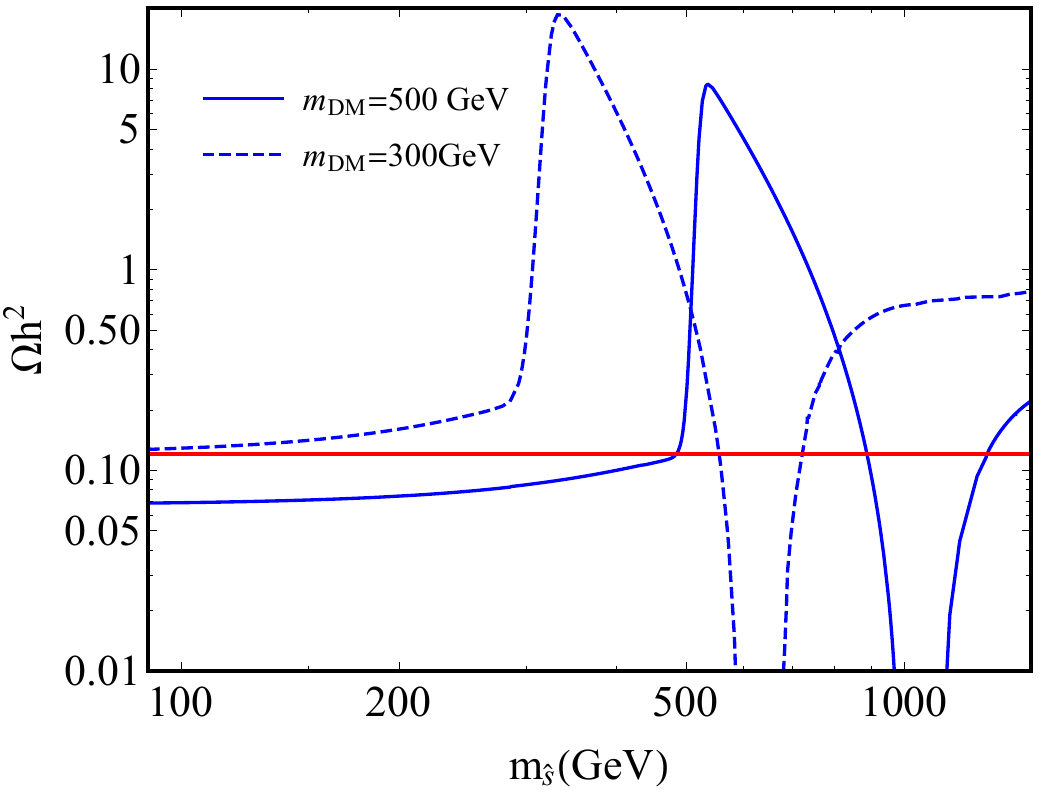}
\caption{$\Omega h^2$ versus the mediator mass $m_{\hat{s}}$ with the solid, dashed lines correspond to $m_{\rm{DM}}$ = 500, 300GeV, respectively.
The red line represents the observed relic abundance $\Omega h^2=0.12$\cite{Planck:2018vyg}. } \label{fig1}
\end{center}
\end{figure}

To analyze the dependence of the DM relic density on free parameters, we show in the Fig.\ref{fig1} $\Omega h^2$ versus the mediator mass $m_{\hat{s}}$ with the blue solid, dashed lines corresponding to $m_{DM}$ = 500, 300GeV, respectively. The horizontal red line represents the observed relic abundance $\Omega h^2=0.12$\cite{Planck:2018vyg}. When $m_{\hat{s}}>m_{DM}$, the processes of containing $\hat{s}$ in the final states are forbidden, so the cross-section suddenly decreases and the relic abundance has a peak. With the $m_{\hat{s}}$ increases, the relic abundance rapidly drops, since there exits a resonant region as the $m_{\hat s}$ approaches to $2m_{DM}$, which we discuss further in Sec. \ref{sec:direct}.
\section{Direct detections}
\label{sec:direct}

\begin{figure}[t]
\begin{center}
\includegraphics[height=6cm,width=8cm]{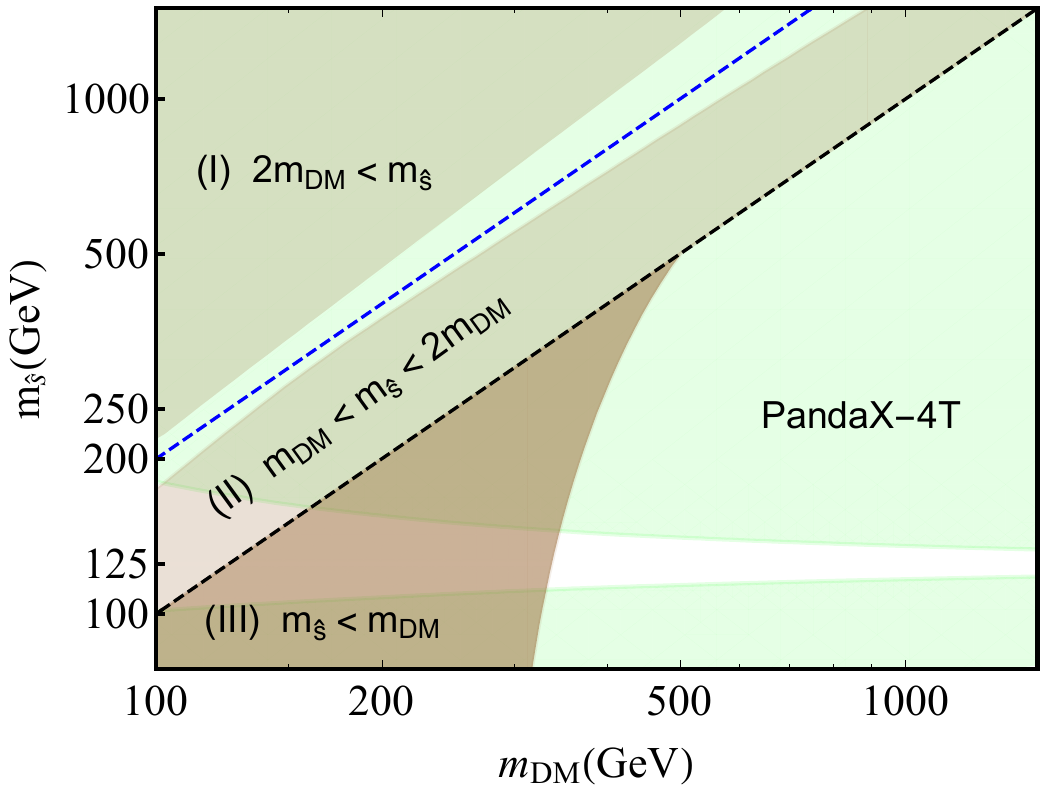}
\caption{The exclusion limit for DM mass $m_{DM}$ vs. $m_{\hat{s}}$. The light-green shaded region is excluded by the PandaX-4T direct detection experiment. The brown shaded regions indicate the relic density $\Omega h^2>0.12$, which divided into three cases: (I) $2m_{DM} < m_{\hat{s}}$, (II) $m_{DM}<m_{\hat{s}}<2m_{DM}$ and (III) $m_{\hat{s}}< m_{DM}$, where the last case (darker brown) has extra annihilation channels $\left(\chi_i \chi_j \to \hat{s} \hat{s}(\hat{s} \hat{h})\right)$. The blue and black dashed lines denote $m_{\hat{s}}=2m_{DM}$ and $m_{\hat{s}}=m_{DM}$, respectively.} \label{fig2}
\end{center}
\end{figure}

Since $\chi_1 $ totally decayed away in the early Universe, the direct detection signal of the pseudo-Dirac DM is actually induced by the scattering of $\chi_2$ off the target nuclei. To calculate the direct detection cross-section, one needs to write down the Wilson coefficients $C_{\rm Wilson}^{}$ for the effective interaction $\bar N N \bar \chi_2^{}  \chi_2^{} $,
\begin{eqnarray}
C_{\rm Wilson}^{} = { m_\chi m_N \sin 2 \alpha \over 4 v_h v_\Phi } \left( {1\over m_{\hat{s}}^2} -{1\over m_{\hat{h}}^2} \right) \sum_{q} f^N_{q},
\end{eqnarray}
where $m_N$ is the mass of the nucleon $(p,n)$, $f^N_q$ is the quark matrix element defined by $\langle N|m_q \bar{q} q|N\rangle=m_N f^N_{q}$.
In this paper, we use the following  nucleon form factors  for light quarks \cite{Abe:2018emu}
\begin{eqnarray}
f^p_u= 0.0153 , \ \ \ f^p_d= 0.0191 , \ \ \ f^p_s= 0.0447 , \nonumber \\
f^n_u= 0.0110 , \ \ \ f^n_d= 0.0273 , \ \ \ f^n_s= 0.0447 , \nonumber  \label{fq}
\end{eqnarray}
where $p$ and $n$ stand for proton and neutron, respectively. Heavy quark matrix element is related to light quark matrix elements via the following formula: $f^N_Q=(2/27) (1-\sum_{q=u,d,s}f^N_q )$~\cite{Freytsis:2010ne}. The $\chi-N$ for the spin-independent scattering cross section  is then
\begin{eqnarray}
\sigma^{\rm SI}=\frac{\mu^2 \sin^2 2\alpha}{16\pi}\left( {m_\chi m_N \over v_h v_\Phi}\right)^2\left(\frac{1}{m^2_{\hat{s}}}-\frac{1}{m^2_{\hat{h}}}\right)^2\left(\frac{2}{9}+\frac{7}{9}\sum_{q=u,d,s}f^N_q\right)^2, \label{direct1}
\end{eqnarray}
where $\mu $ is the reduced mass of the DM-nucleon system, $\mu = m_\chi m_N /(m_\chi+m_N)$.

We show in Fig.~\ref{fig2} the available parameter space in the $m_{\rm DM}-m_{\hat s}$ plane by considering the exclusion limit put by the  DM direct detection experiment PandaX-4T~\cite{PandaX-4T:2021bab} as well as the constraint put by the observed DM relic abundance $\Omega h^2=0.12$~\cite{Planck:2018vyg,ParticleDataGroup:2020ssz}. When making the plot we set $\upsilon_\Phi=1000~{\rm GeV}, m_{\hat a}=100~{\rm GeV}, \cos\alpha=0.9$, and assume $m_{\rm DM} > m_{\hat h}/2$ so as to avoid the bound of Higgs to invisible decay. The light green regime is excluded by the PandaX-4T and the light brown regime is excluded by the observed relic abundance, which is divided into three parts: (I) $2 m_{\rm DM} <m_{\hat s } $, (II) $m_{\rm DM} < m_{\hat s} < 2 m_{\rm DM}$ and (III) $m_{\hat s } < m_{\rm DM}$. For the parameter between regime (\uppercase\expandafter{\romannumeral1}) and (\uppercase\expandafter{\romannumeral2}), the total DM annihilation cross section is resonantly enhanced, resulting in the relatively small relic abundance that satisfies constraint. For the regime (\uppercase\expandafter{\romannumeral3}), the new annihilation channel $\bar \chi_i \chi_j \to \hat s \hat s(\hat h \hat s)$ are kinematically allowed, which enhance the total annihilation cross section. Notice that the regime with  $m_{\hat s}\sim 125~{\rm GeV}$ is always allowed by the direct detection, due to the cancellation between CP-even mediators in the direct detection cross section, as can be seen in Eq.(\ref{direct1}).

\section{The effective number of neutrinos}
\label{sec:effective}

The effective number of neutrinos ($N_{\rm eff}$) with the default value $3$ is an important cosmological parameter for probing the thermal history of the early Universe.  The Planck collaboration reports the precision measurements of $N_{\rm eff}^{} $, which has $N_{\rm eff} =2.99\pm0.34$ at $95\%$ CL~\cite{Planck:2018vyg} in the framework of $\Lambda$CDM.  The theoretical value  for $N_{\rm eff}^{} $  comes from solving Boltzmann equations for energy densities of active neutrinos and photon,  and the $N_{\rm eff}$ is defined as~\cite{Shvartsman:1969mm,Steigman:1977kc}
\begin{eqnarray}
N_{\rm eff}=\frac{8}{7}\left(\frac{11}{4}\right)^{4/3}\left(\frac{\rho_\nu}{\rho_\gamma}\right)=3\left(\frac{11}{4}\right)^{4/3}\left(\frac{T_\nu}{T_\gamma}\right)^{4}, \label{Neff}
\end{eqnarray}
where $\rho_\nu$ and $\rho_\gamma$ are energy densities of active neutrinos and photon with $T_\nu$ and $T_\gamma$ being their corresponding temperatures, respectively.
In order to study the impact of beyond SM physics, we define the amount of the change,  $\Delta{N_{\rm eff}}= N_{\rm eff}-N^{\rm SM}_{\rm eff}$, where  $N^{\rm SM}_{\rm eff}$=3.045 is the prediction of the minimal SM~\cite{deSalas:2016ztq,Mangano:2005cc,EscuderoAbenza:2020cmq}.

In our model, the excited DM state $\chi_1$ may decay into radiations during or after the freeze-out of active neutrinos, resulting in a modification to the $N_{\rm eff}^{}$. How large is $\Delta N_{\rm eff}$  induced by $\chi_1$ and can it be a signal of excited DM state are questions need to clarified. To do so, we start by reviewing the Boltzmann equation that governs the evolution of a given species~\cite{Starkman:1993ik,Escudero:2018mvt,Ng:1993nh,Hannestad:1995rs}
\begin{eqnarray}
&&\frac{d \rho}{dt}+3H(\rho+p)=\int g E\frac{d^3 p}{(2\pi)^3}\mathcal{C}[f],\label{Boltzmann}
\end{eqnarray}
where $\rho$ and $p$ are the  energy, and pressure densities of the given species, $g$ represents the  internal degrees of freedom. In our model, $\chi_1$ is smashed via a three-body decay  $\chi_1 \to \chi_2 + 2 \gamma$, and its collision term reads\cite{Kawasaki:1992kg,Dolgov:2002wy,Hasegawa:2019jsa,Kawasaki:2000en}
\begin{eqnarray}
\begin{aligned}
\mathcal{C}[f]&=-\frac{1}{2E_1}\int\prod_i d\Pi_i(2\pi)^4\delta^4(p_1-p_2-p_3-p_4)\Big[|\mathcal{M}|^2 f_1(1\pm f_2)(1\pm f_3)(1\pm f_4)\\
&-(1\pm f_1)f_2 f_3 f_4\Big], \label{collision}
\end{aligned}
\end{eqnarray}
where $f_1$, $f_2$, $f_3$ and $f_4$ are the distribution functions of $\chi_1$, $\chi_2$ and two photons respectively. $d\Pi_i={1 \over (2\pi)^3}{d^3p_i \over 2E_i}$ represents the phase space of $i$th particle.

\begin{figure}[t]
\begin{center}
\includegraphics[height=6cm,width=8cm]{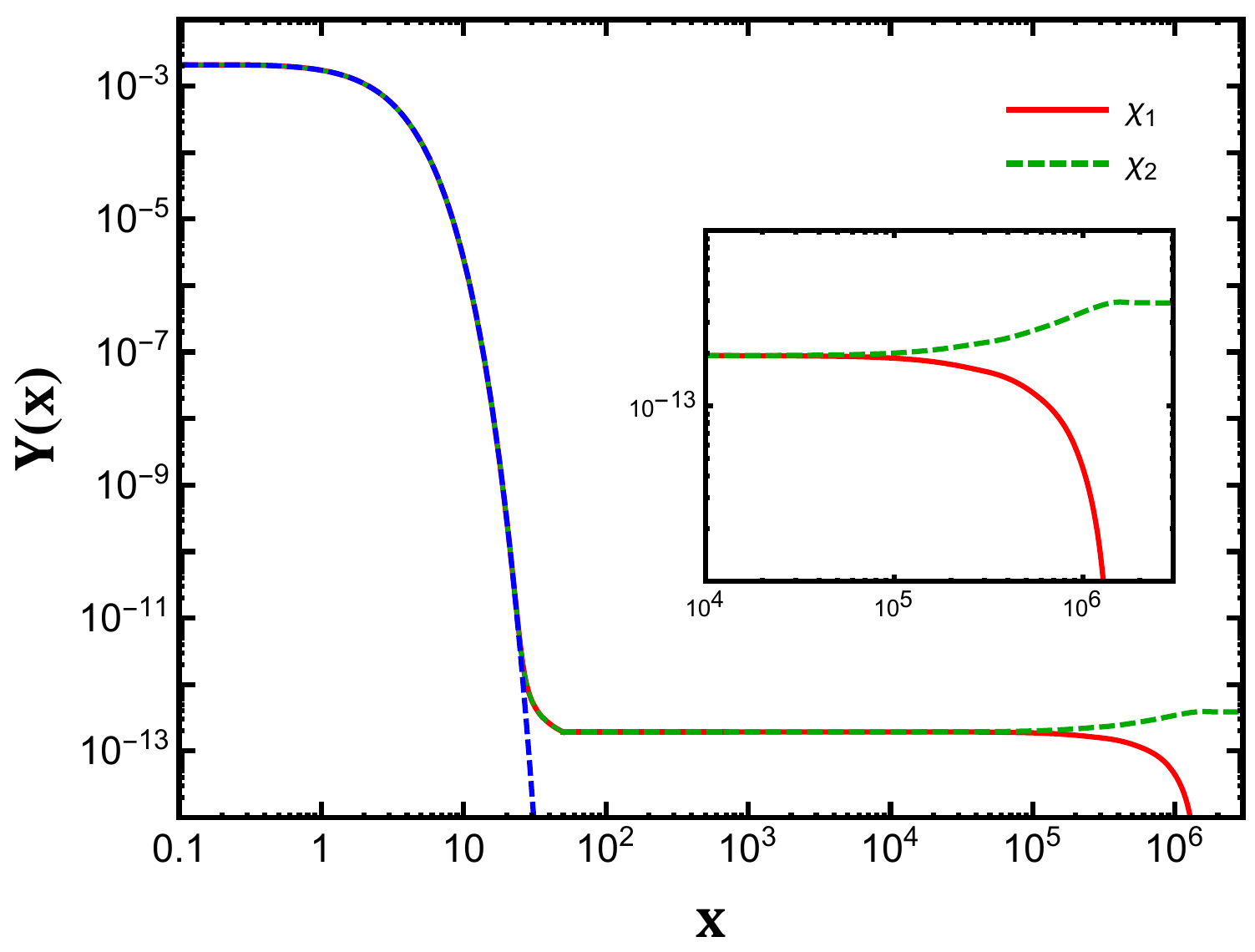}
\caption{Evolution of comoving number density $Y=n/s$ with $x=m/T$. The red solid line represents species $\chi_1$, where the corresponding $Y$ become fixed after freeze-out and it begins to decrease when $\chi_1$ decay. The green dashed line denotes species $\chi_2$, which increases by the decay of $\chi_1$.} \label{fig4}
\end{center}
\end{figure}

\begin{figure}[t]
\begin{center}
\includegraphics[height=6cm,width=9cm]{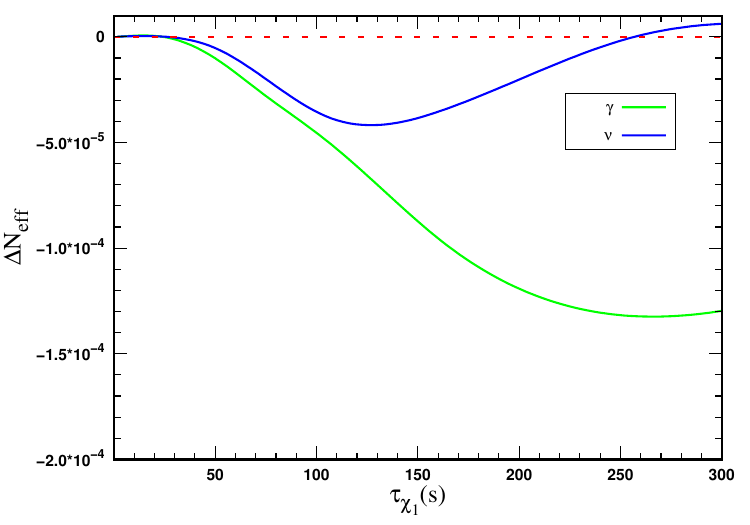}
\caption{$\Delta N_{\rm eff}$ as a function of the lifetime $\tau_{\chi_1}$ that can decay into photons and neutrinos through $\chi_1\rightarrow\chi_2+2\gamma$ and $\chi_1\rightarrow\chi_2+4\nu$ processes, respectively. The green solid line means that the decay product is photons and the blue line describes that decay into neutrinos. The red dashed line is $\Delta N_{\rm eff}=0$.} \label{fig5}
\end{center}
\end{figure}

Given  Eqs.(\ref{Boltzmann}) and (\ref{collision}), we can numerically solve  the energy density of a massive decay particle,  following the method developed in Refs.~\cite{Blackadder:2014wpa,Escudero:2019gzq,Luo:2020sho}.  Since we are only interested in the case of  $\chi_1$ that is decoupled from the thermal bath, we can take $p\sim 0$ and ignore  the term proportional to $f_{\chi1}$, so the Boltzmann equation can be simplified as
\begin{eqnarray}
\frac{d\rho_{\chi 1} }{dt}+3H\rho_{\chi1}=-\Gamma_0 \rho_{\chi 1} \; .\label{Boltzmann1}
\end{eqnarray}
The $\Gamma_0 (\chi_1 \to \chi_2 + 2 \gamma )$ can be calculated in the rest frame of $\chi_1$,
\begin{eqnarray}
\begin{aligned}
\Gamma_0&=\frac{y^2\cos^2(2\theta)\sin^2(2\alpha)C^2_{eff}}{16\pi^3}\frac{m^7_1}{240}\bigg[
(1-x^2)(1-5x-14x^2-145x^3-94x^4-145x^5\\
&-14x^6-5x^7+x^8)-120x^3(1+x+3x^2+x^3+x^4)\log{(x)}\left(\frac{1}{m^4_{\hat{s}}}+\frac{1}{m^4_{\hat{h}}}-{2\over m^2_{\hat{h}} m^2_{\hat{s}}}\right)\bigg],
\end{aligned}
\end{eqnarray}
where $x=m_2/m_1$, $C_{eff} =\frac{\alpha_{EM}}{2\pi}Q^2_t \frac{g}{2 m_w} $, and more  details of the calculation are given in the appendix \ref{appendix B}.

Solving the differential equation in Eq.(\ref{Boltzmann1}), we obtain,
\begin{eqnarray}
\rho_1(a) = \rho_* e^{-\Gamma_0^{} [t (a) -t(a_*)]} \left( a_* \over a \right)^3,
\end{eqnarray}
where $\rho_* =\rho(a_*)$ is the energy density of $\chi_1$ with $a_*$ the scale factor when $\chi_1$ starts to decay.

We show in the Fig.\ref{fig4}  $Y=n/s$ as the function of $x=m/T$, where $n$ represents the number density, and $s$ represents the entropy density of the Universe.   We set $m_{DM}=300\rm GeV, \Delta m=0.3\rm GeV$ and $m_{\hat{a}}=100 \rm GeV$. The blue dashed line represents the evolution of particles at equilibrium. The red solid line and the green dashed line represents the evolution of $\chi_1$ and $\chi_2$, respectively. At the beginning, the $\chi_{1,2}$ are both in thermal equilibrium. As the temperature decreases, the $\chi_{1,2}$ gradually depart from thermal equilibrium and freeze out. Then $\chi_1$ begins to decay via the process $\chi_1\rightarrow\chi_2+2\gamma$. Therefore, $Y_{\chi_1}$ (red solid) drops until $\chi_1$ totally decays away. In the meanwhile, the $Y_{\chi_2}$ (green dashed) increases during the $\chi_1$ decay and then the curve tends to flat.

Now we can calculate the $\Delta N_{\rm eff}$ induced by the decay $\chi_1\rightarrow\chi_2+2\gamma$. Since the energy transferred by $\chi_1$ is not totally deposited into radiation, we need to introduce the parameter $\varepsilon=(m_1-m_2)/m_1$~\cite{Blackadder:2014wpa}, which denotes the fraction of the energy of the $\chi_1$ that has been transferred to $\gamma$.
We neglect the influence of chemical potential and write down the temperature evolution equations from Eq.(\ref{Boltzmann}) and Eq.(\ref{Boltzmann1}) by assuming that the temperature of three generation active neutrinos are the same,
\begin{eqnarray}
&&\frac{dT_\gamma}{dt}=-\frac{4H\rho_\gamma+3H(\rho_e+p_e)+\frac{\delta \rho_{\nu_e}}{\delta t}+2\frac{\delta \rho_{\nu_\mu}}{\delta t}-\varepsilon \Gamma_0 \rho_1}{\frac{\partial \rho_\gamma}{\partial T_\gamma}+\frac{\partial \rho_e}{\partial T_\gamma}},  \label{2gamma_a}\\
&&\frac{dT_\nu}{dt}=-HT_\nu+\frac{\frac{\delta\rho_{\nu e}}{\delta_t}+2\frac{\delta\rho_{\nu\mu}}{\delta t}}{3\frac{\partial \rho_\nu}{\partial T_\nu}}, \label{2gamma_b}
\end{eqnarray}
where $\frac{\delta \rho_{\nu_e}}{\delta t}$ and $\frac{\delta \rho_{\nu_\mu}}{\delta t}$ are  collision terms induced by neutrino interactions calculated in the Refs.\cite{Dolgov:2002wy,Hannestad:1995rs,Escudero:2018mvt}. Given $T_\nu$ and $T_\gamma$, one can estimate $\Delta N_{\rm eff}^{}$ numerically.

It should be mentioned that Eqs.~(\ref{2gamma_a}) and (\ref{2gamma_b}) only estimate the effect of  $\chi_1 \to \chi_2 +2\gamma$. Actually, it allows the decay process $\chi_1\rightarrow \chi_2+4\nu$ if there is an additional gauge boson $Z^\prime$ coupled to scalar singlet. For this case, the energy of $\chi_1$ is transferred into neutrinos, which is then transferred into positron and electron, thus affecting the temperature evolution of the photon\cite{Boyarsky:2021yoh}. In this case, the temperature evolution equations become:
\begin{eqnarray}
&&\frac{dT_\gamma}{dt}=-\frac{4H\rho_\gamma+3H(\rho_e+p_e)+\frac{\delta \rho_{\nu_e}}{\delta t}+2\frac{\delta \rho_{\nu_\mu}}{\delta t}-\varepsilon \xi_{EM} \Gamma_0 \rho_1}{\frac{\partial \rho_\gamma}{\partial T_\gamma}+\frac{\partial \rho_e}{\partial T_\gamma}}, \label{gama} \\
&&\frac{dT_\nu}{dt}=-HT_\nu+\frac{\frac{\delta\rho_{\nu e}}{\delta_t}+2\frac{\delta\rho_{\nu\mu}}{\delta t}+\varepsilon (1-\xi_{EM}) \Gamma_0 \rho_1}{3\frac{\partial \rho_\nu}{\partial T_\nu}}, \label{nu}
\end{eqnarray}
where $\xi_{EM}$ is energy fraction that the neutrinos produced by decay process inject into electromagnetic plasma~\cite{Boyarsky:2021yoh}.
\begin{eqnarray}
\xi_{EM}=\sum^N_{k=0}(\frac{1}{2}P_3+P_2)(P_1+\frac{1}{2}P_3)^k.\label{fraction}
\end{eqnarray}
Here $P_1$, $P_2$ and $P_3$ are the average probabilities of the following processes: $\nu_{non-eq}+\nu_{therm}\rightarrow\nu_{non-eq}+\nu_{non-eq}$, $\nu_{non-eq}+\bar{\nu}_{therm}\rightarrow e^{+}+e^{-}$ and $\nu_{non-eq}+e^{\pm}\rightarrow\nu_{non-eq}+e^{\pm}$, respectively, where `non-eq'  and `therm' denote neutrinos from non-equilibrium and thermal energies. Following Ref.~\cite{Sabti:2020yrt}, one has $P_1\approx0.76$, $P_2\approx0.05$, $P_3\approx0.19$, and $N=\log_2({E_{inj}/3.15T})$. $E_{inj}$ is the injected neutrino energy. Through the temperature evolution equations (\ref{gama}-\ref{nu}), we can obtain the temperature of photons and neutrinos in the CMB epoch.

The Fig.\ref{fig5} shows $\Delta N_{\rm eff}$ as the function of the lifetime $\tau_{\chi_1}$. The red dashed line is $\Delta N_{\rm eff}=0$. The green solid line represents the contribution of the process  $\chi_1\rightarrow \chi_2+2\gamma$ to the $N_{\rm eff}$. Since the energy of $\chi$ is partially deposited into the photon and the energy density of neutrinos unchanges, the $N_{\rm eff}^{}$ will decrease compared with the SM prediction resulting in a negative $\Delta N_{\rm eff}^{} $ which is of the order $10^{-4}$.
The blue solid line is the contribution of the decay process $\chi_1\rightarrow \chi_2+4\nu$, which not only increases the temperature of neutrinos but also distributes energy to the electromagnetic plasma. We find that when $E_{inj}=300 \rm MeV$, the $\xi_{EM}$ increases as the temperature drops by Eq.(\ref{fraction}). And the neutrinos produced by decay process transfer a significant amount of their energy to the EM plasma. As temperature decreases further, the weak reactions go out of equilibrium and the energy transfer efficiency is getting lower and lower. Therefore, the $\Delta N_{\rm eff}$ first decreases and then increases until it is bigger than 0 due to the change of the energy allocated to the EM plasma.

\section{Conclusion and Summary}
\label{sec:Summary}
In this paper, we work in the framework of pseudo-Dirac DM with Higgs portal by calculating the available parameter space of the model and studying the impact of DM excited state on the effective number of neutrino species. The pseudo-Dirac fermion $\chi$ can split into two non-degenerate Majorana fermion mass eigenstates $\chi_1$ (heavy) and $\chi_2$ (light), where $\chi_2$ can be a candidate for cold DM. Using the DM relic abundance $\Omega h^2=0.12$ and the limit of the PandaX-4T direct detection experiment, we find the feasibility parameter space of the model. Next, we calculated the effective number of neutrinos. We found that cold DM decay, no matter whether the decay products are photons or neutrinos, the impact on $\Delta N_{\rm eff}$ is relatively small, and the maximum is of the order $10^{-4}$.

\vspace{1cm}
{\bf Acknowledgments}
This work was supported by the National Natural Science Foundation of China under grant No. 11775025, No. 12175027 and the Fundamental Research Funds for the Central Universities under grant No. 2017NT17.  The authors thank Siyu Jiang for helpful discussion.

\begin{appendix}
\section{The cross section of $\sigma_{ij}$}\label{appendix A}
In this section we calculate two-body annihilation processes, $a(p_a)+b(p_b)\rightarrow3(p_3)+4(p_4)$. Using the momentum-energy relationship, in the center of mass frame we can get
\begin{eqnarray}
\begin{aligned}
|\vec{p}_a|&=|\vec{p}_b|=\sqrt{\frac{\left(s-(m_a+m_b)^2\right) \left(s-(m_a-m_b)^2\right)}{4 s}},\\
|\vec{p}_3|&=|\vec{p}_4|=\sqrt{\frac{\left(s-(m_3+m_4)^2\right) \left(s-(m_3-m_4)^2\right)}{4 s}}.
\end{aligned}
\end{eqnarray}
Therefore, the differential annihilation cross section can be written as
\begin{eqnarray}
\begin{aligned}
{d\sigma\over d\Omega}={1\over 64 \pi^2 s}{|\vec{p}_3|\over|\vec{p}_a|}\sum_{spins}|\mathcal{M}|^2.
\end{aligned}
\end{eqnarray}
If final states are identical particles, we have the symmetry factor ${1\over 2!}$.
\subsection{$\chi_2 \bar{\chi}_2$ annihilation processes}
According to the type of the final particles, there are four cases(fermion, vector, same scalar and  different scalar).
\subsubsection{ $\chi_2 \bar{\chi}_2\rightarrow \bar{f}f$ \rm (fermion case)}
By using the Eq.~(\ref{coupling}) and the SM interactions, the coupling are given by,
\begin{eqnarray}
\begin{aligned}
&\mathcal{L}_{ \chi_2 \chi_2 \hat h(\hat s)}={y\sin{2\theta}\over 2\sqrt{2}}\bar{\hat{\chi}}_{2}\left(\sin{\alpha} \hat{h}+\cos{\alpha} \hat{s}\right)\hat{\chi}_{2},\\
&\mathcal{L}_{ff \hat h(\hat s)}=-{g m_{f}\over 2 m_w}\bar{f}f(\cos{\alpha}\hat{h}-\sin{\alpha} \hat{s}).
\end{aligned}
\end{eqnarray}
For $s$-channel with the SM Higgs($h$) and CP-even scalar($s$) exchange, the squared amplitude reads
\begin{eqnarray}
\begin{aligned}
|\mathcal{M}|^2=&\sum_{spins}|\bar{u}(p_3)C_{ff\hat h}\nu(p_4){i\over k^2-m_{\hat{h}}^2}\bar{\nu}(p_b)C_{\chi_2\chi_2 \hat h}u(p_a)\\
&+\bar{u}(p_3)C_{ff \hat s}\nu(p_4){i\over k^2-m_{\hat{s}}^2}\bar{\nu}(p_b)C_{\chi_2\chi_2 \hat s}u(p_a)|^2.
\end{aligned}
\end{eqnarray}
Since $k^2=(p_a+p_b)^2=s$, $m_a=m_b=m_2$ and $m_3=m_4=m_f$, one can simplify the annihilation cross section and then get the final result,

\begin{eqnarray}
\sigma=\frac{g^2 m_f^2 y^2 \sin ^2(2 \alpha ) \sin ^2(2 \theta ) \sqrt{s-4 m_2^2} \left(m_{\hat{h}}^2-m_{\hat{s}}^2\right)^2 \left(s-4 m_{f}^2\right)^{3/2}}{2048 \pi  m_w^2 s \left(m_{\hat{h}}^2-s\right)^2 \left(m_{\hat{s}}^2-s\right)^2}, \label{22ff}
\end{eqnarray}
where $s$ is the usual Mandelstam variable, $m_f$ is a leptons or quarks mass. Free parameters $\theta, \alpha, y$ are all defined in section \ref{sec:Model}. $g$ is the weak gauge coupling constant and $m_w$ is the mass of $W$ boson in the SM.

\subsubsection{$\chi_2 \bar{ \chi}_2\rightarrow VV(ZZ,W^{+}W^{-})$\rm (vector case)}
In this case the coupling are as follow,
\begin{eqnarray}
\begin{aligned}
&\mathcal{L}_{WW \hat h(\hat s)}=g m_w W^{+}_\mu W^{-\mu}(\cos{\alpha}\hat{h}-\sin{\alpha} \hat{s}),\\
&\mathcal{L}_{ZZ \hat h(\hat s)}={g m_w\over 2c_w^2}Z_\mu Z^{\mu}(\cos{\alpha}\hat{h}-\sin{\alpha} \hat{s}).
\end{aligned}
\end{eqnarray}
For $s$-channel with the SM Higgs($\hat h$) and CP-even scalar($\hat s$) exchange, the squared amplitude reads
\begin{eqnarray}
\begin{aligned}
|\mathcal{M}|^2=&\sum_{\epsilon}\sum_{spins}|\bar{\nu}(p_b)C_{\chi_2\chi_2 \hat h}u(p_a){i\over k^2-m_{\hat{h}}^2}C_{WW\hat h}g^{\mu\nu}\varepsilon^{*}_{\mu}(p_3)\varepsilon^{*}_{\nu}(p_4)\\
&+\bar{\nu}(p_b)C_{\chi_2\chi_2 \hat s}u(p_a){i\over k^2-m_{\hat{s}}^2}C_{WW\hat s}g^{\mu\nu}\varepsilon^{*}_{\mu}(p_3)\varepsilon^{*}_{\nu}(p_4)|^2,
\end{aligned}
\end{eqnarray}
where $m_a=m_b=m_2$ and $m_3=m_4=m_V$. When summing over vector polarization, we can get
\begin{eqnarray}
\sigma =&&\frac{g^2 m_w^2 y^2 \sin ^2(2 \alpha ) \sin ^2(2 \theta ) \sqrt{(s-4 m_2^2)(s-4 m_V^2)} \left(m_{\hat{h}}^2-m_{\hat{s}}^2\right)^2 }{4096 \pi  m_V^4 s \left(m_{\hat{h}}^2-s\right)^2 \left(m_{\hat{s}}^2-s\right)^2} \nonumber \\
&&\times \left(12 m_V^4-4 m_V^2 s+s^2\right).
\label{22VV}
\end{eqnarray}
In case of the final states being $ZZ$, the expressions for $\sigma$ needs to be multiplied by ${1\over 2 c^4_w}$ and $m_V=m_Z$. If the final states are $W^{+}W^{-}$, $m_V=m_w$.
\subsubsection{$\chi_2 \bar{\chi}_2\rightarrow SS$($\hat h \hat h, \hat s \hat s $ and $\hat a \hat a$)\rm (same scalar case)}
The squared amplitude reads
\begin{eqnarray}
\begin{aligned}
|\mathcal{M}|^2&=|\bar{\nu}(p_b)C_{\chi_2\chi_2 \hat h}u(p_a){i\over k^2-m_{\hat{h}}^2}C_{SS\hat h}+\bar{\nu}(p_b)C_{\chi_2\chi_2 \hat h}u(p_a){i\over k^2-m_{\hat{s}}^2}C_{SS\hat s}\\
&+\bar{\nu}(p_b)C_{\chi_2\chi_2 S}{\slashed{k_1}+m_2 \over k_1^2-m_2^2}C_{\chi_2\chi_2 S}u(p_a)+\bar{\nu}(p_b)C_{\chi_2\chi_2 S}{\slashed{k_2}+m_2 \over k_2^2-m_2^2}C_{\chi_2\chi_2 S}u(p_a)|^2,
\end{aligned}
\end{eqnarray}
where $k_1=p_a-p_3$, $k_2=p_a-p_4$, $m_a=m_b=m_2$ and $m_3=m_4=m_S$.
For $s$-channel with SM Higgs($\hat h$), CP-even scalar($\hat s$) and  $t,u$-channel with $\chi_{1,2}$ exchange, we obtain
\begin{eqnarray}
\begin{aligned}
\sigma=&{y^2 \sin ^2(2 \theta ) \sqrt{s-4 m_S^2} \over 2048 \pi  s \sqrt{s-4 m_2^2}}  \Bigg\{\frac{96 C_{SS\hat h} C_{SS\hat s} \sin (\alpha ) \cos (\alpha ) \left(4 m_2^2-s\right)}{\left(m_{\hat{h}}^2-s\right) \left(s-m_{\hat{s}}^2\right)}\\
&+16 \sin ^2(\alpha ) \Bigg(\frac{9 C_{SS\hat h}^2 \left(s-4 m_2^2\right)}{\left(m_{\hat{h}}^2-s\right)^2}+\frac{\sqrt{2} m_2 C_{SS \hat s} y \cos (\alpha ) \sin (2 \theta )}{m_{\hat{s}}^2-s}\Bigg)\\
&-\frac{48 m_2 C_{SS\hat h} y \sin ^3(\alpha ) \sin (2 \theta ) \left(-8 m_2^2+2 m_S^2+s\right) \arctan\left(\frac{\sqrt{s-4 m_2^2} \sqrt{4 m_S^2-s}}{s-2 m_S^2}\right)}{\sqrt{s-4 m_2^2} \left(s-m_{\hat{h}}^2\right) \sqrt{2 m_S^2-\frac{s}{2}}}\\
&-\frac{8 m_2 C_{SS\hat s} y \sin (2 \alpha ) \sin (\alpha ) \sin (2 \theta ) \left(-8 m_2^2+2m_S^2+s\right) \arctan\left(\frac{\sqrt{s-4 m_2^2} \sqrt{4 m_S^2-s}}{s-2 m_S^2}\right)}{\sqrt{s-4 m_2^2} \left(s-m_{\hat{s}}^2\right) \sqrt{2 m_S^2-\frac{s}{2}}}\\
&+\frac{16 C_{SS\hat s}^2 \cos ^2(\alpha ) \left(s-4 m_2^2\right)}{\left(m_{\hat{s}}^2-s\right)^2}+\frac{y^2 \sin ^4(\alpha ) \sin ^2(2 \theta ) \left(m_S^2-4 m_2^2\right)^2}{m_2^2 \left(4 m_S^2-s\right)-m_S^4}\\
&-\frac{2 y^2 \sin ^4(\alpha ) \sin ^2(2 \theta ) \left(32 m_2^4+16 m_2^2 \left(m_S^2-s\right)-6 m_S^4+4 m_S^2 s-s^2\right) \arctan\left(\frac{\sqrt{4 m_2^2-s} \sqrt{s-4 m_S^2}}{s-2 m_S^2}\right)}{\sqrt{4 m_2^2-s} \sqrt{s-4 m_S^2} \left(s-2 m_S^2\right)}\\
&+\frac{48 \sqrt{2} m_2 C_{SS\hat h} y \sin ^3(\alpha ) \sin (2 \theta )}{m_{\hat{h}}^2-s}-2 y^2 \sin ^4(\alpha ) \sin ^2(2 \theta )
\Bigg\}. \label{22sss}
\end{aligned}
\end{eqnarray}

The above equation is merely a dominate term in this case, some terms with $\cos^2{2 \theta}$ are neglected. When the final states are $\hat{h}\hat{h}$ and $\hat{s}\hat{s}$, $m_S= m_{\hat{h}}$ and $m_S= m_{\hat{s}}$ in the expression of $\sigma$, respectively. The coefficients of the scalar interaction $C_{SS \hat s}$ and $C_{SS\hat h}$ are obtained by the potential $V\left(H,\Phi\right)$ in Eq.~(\ref{eq:vscalar}).
Since the portal interactions of $\chi_{1}$ are the same as $\chi_{2}$, the above all processes are also applied to $\chi_1\chi_1$ annihilation processes, where one only needs to replace $m_1$ with $m_2$ for calculating $\sigma$.
\subsubsection{$\chi_2 \bar{\chi}_2\rightarrow \hat{s}\hat{h}$\rm (different scalar case)}
The squared amplitude is similar to the previous one. For $m_a=m_b=m_2$, $m_3=m_{\hat{h}}$ and $m_4=m_{\hat{s}}$, we obtain
\begin{eqnarray}
\begin{aligned}
\sigma=&\frac{\sqrt{\frac{\left(s-(m_{\hat{h}}+m_{\hat{s}})^2\right) \left(s-(m_{\hat{h}}-m_{\hat{s}})^2\right)}{4 s}}}{128 \pi  s \sqrt{\left({s\over 4}-m_2^2\right)}}y^4 \sin ^2(2 \alpha ) \sin ^4(2 \theta ) \Bigg\{\frac{ s p%\sqrt{1-\frac{4 m_2^2}{s}}
\left(8 m_2^2-m_{\hat{h}}^2+m_{\hat{s}}^2\right) \left(m_{\hat{h}}^2-3 m_{\hat{s}}^2+s\right)}{128 \left(s-4 m_2^2\right) q%\sqrt{\frac{m_{\hat{h}}^4-2 m_{\hat{h}}^2 \left(m_{\hat{s}}^2+s\right)+\left(m_{\hat{s}}^2-s\right)^2}{s}}
\left(2 p q%\sqrt{1-\frac{4 m_2^2}{s}} \sqrt{\frac{m_{\hat{h}}^4-2 m_{\hat{h}}^2 \left(m_{\hat{s}}^2+s\right)+\left(m_{\hat{s}}^2-s\right)^2}{s}}
+m_{\hat{h}}^2-3 m_{\hat{s}}^2+s\right)}\\
&+\frac{  s p%\sqrt{1-\frac{4 m_2^2}{s}}
\left(8 m_2^2+m_{\hat{h}}^2-m_{\hat{s}}^2\right) \left(-3 m_{\hat{h}}^2+m_{\hat{s}}^2+s\right)}{128 \left(s-4 m_2^2\right)q %\sqrt{\frac{m_{\hat{h}}^4-2 m_{\hat{h}}^2 \left(m_{\hat{s}}^2+s\right)+\left(m_{\hat{s}}^2-s\right)^2}{s}}
\left(2 p q %\sqrt{1-\frac{4 m_2^2}{s}} \sqrt{\frac{m_{\hat{h}}^4-2 m_{\hat{h}}^2 \left(m_{\hat{s}}^2+s\right)+\left(m_{\hat{s}}^2-s\right)^2}{s}}
-3 m_{\hat{h}}^2+m_{\hat{s}}^2+s\right)}\\
&+\frac{- \left(64 m_2^4+16 m_2^2 \left(m_{\hat{h}}^2-m_{\hat{s}}^2-s\right)+m_{\hat{h}}^4-2 m_{\hat{h}}^2 \left(m_{\hat{s}}^2-2 s\right)+m_{\hat{s}}^4-s^2\right)}{256 p q%\sqrt{1-\frac{4 m_2^2}{s}} \sqrt{\frac{m_{\hat{h}}^4-2 m_{\hat{h}}^2 \left(m_{\hat{s}}^2+s\right)+\left(m_{\hat{s}}^2-s\right)^2}{s}}
\left(-2 p q %\sqrt{1-\frac{4 m_2^2}{s}} \sqrt{\frac{m_{\hat{h}}^4-2 m_{\hat{h}}^2 \left(m_{\hat{s}}^2+s\right)+\left(m_{\hat{s}}^2-s\right)^2}{s}}
-3 m_{\hat{h}}^2+m_{\hat{s}}^2+s\right)}\\
&+\frac{ \left(64 m_2^4+16 m_2^2 \left(m_{\hat{h}}^2-m_{\hat{s}}^2-s\right)+m_{\hat{h}}^4-2 m_{\hat{h}}^2 \left(m_{\hat{s}}^2-2 s\right)+m_{\hat{s}}^4-s^2\right)}{256 p q %\sqrt{1-\frac{4 m_2^2}{s}} \sqrt{\frac{m_{\hat{h}}^4-2 m_{\hat{h}}^2 \left(m_{\hat{s}}^2+s\right)+\left(m_{\hat{s}}^2-s\right)^2}{s}}
\left(2 p q %\sqrt{1-\frac{4 m_2^2}{s}} \sqrt{\frac{m_{\hat{h}}^4-2 m_{\hat{h}}^2 \left(m_{\hat{s}}^2+s\right)+\left(m_{\hat{s}}^2-s\right)^2}{s}}
-3 m_{\hat{h}}^2+m_{\hat{s}}^2+s\right)}\Bigg\}.
\end{aligned}
\end{eqnarray}
For simplicity, we define $p\equiv \sqrt{1-\frac{4 m_2^2}{s}}, q\equiv \sqrt{\frac{m_{\hat{h}}^4-2 m_{\hat{h}}^2 \left(m_{\hat{s}}^2+s\right)+\left(m_{\hat{s}}^2-s\right)^2}{s}}$.

\subsection{$\chi_2 \bar{\chi}_1$ annihilation processes}

Because squared amplitude is similar to the corresponding process in the annihilation of $\chi_2 \bar{\chi}_2$, the following will not list them one by one. The difference is that $m_a=m_1$ and $m_b=m_2$.

\subsubsection{$ \chi_2 \bar{\chi}_1\rightarrow \bar{f}f$\rm (fermion case)}
This process involves the s-channel SM Higgs($\hat h$) and CP-even scalar($\hat s$) exchange:
\begin{eqnarray}
\begin{aligned}
\sigma&=\frac{g^2 m_f^2 y^2 \sin ^2(2 \alpha ) \cos ^2(2 \theta ) \left(m_{\hat{h}}^2-m_{\hat{s}}^2\right)^2 \left(s-4 m_f^2\right)^{3/2} \left(-m_1^2+2 m_1 m_2-m_2^2+s\right)}{512 \pi  m_w^2 s \left(m_{\hat{h}}^2-s\right)^2 \left(m_{\hat{s}}^2-s\right)^2 \sqrt{\frac{\left(s-(m_1-m_2)^2\right) \left(s-(m_1+m_2)^2\right)}{s}}}.
\label{12ff}
\end{aligned}
\end{eqnarray}

\subsubsection{$ \chi_2 \bar{ \chi}_1\rightarrow VV(ZZ,W^{+}W^{-})$\rm (vector scalar case)}
This process involves the s-channel SM Higgs($\hat h$) and CP-even scalar($\hat s$) exchange:
\begin{eqnarray}
\begin{aligned}
\sigma&=\frac{g^2 m_w^2 y^2  \sin ^2(2 \alpha )  \cos ^2(2 \theta ) \left(m_{\hat{h}}^2-m_{\hat{s}}^2\right)^2 \sqrt{s-4 m_V^2} \left(12 m_V^4-4 m_V^2 s+s^2\right) \left(-2 m_1^2+2 m_1 m_2+s\right)}{1024 \pi  m_V^4 s \sqrt{s-4 m_1^2} \left(m_{\hat{h}}^2-s\right)^2 \left(m_{\hat{s}}^2-s\right)^2}.\\ \label{12vv}
\end{aligned}
\end{eqnarray}

This is the same as $ \chi_2 \bar{ \chi}_2\rightarrow VV$. When the final states are $ZZ$, the expressions for $\sigma$ needs to be multiplied by ${1\over 2 c^4_w}$ and $m_V=m_Z$. The final states are $W^{+}W^{-}$, $m_V=m_w$.
\subsubsection{$ \chi_2 \bar{\chi}_1\rightarrow SS$($\hat h \hat h, \hat s \hat s $ and $\hat a \hat a$)\rm (same scalar case)}
\begin{eqnarray}
\begin{aligned}
\sigma&=\frac{C_{SS\hat s}^2 y^2 \cos ^2\alpha  \cos ^2(2 \theta ) \sqrt{s-4 m_S^2} \left(-m_1^2+2 m_1 m_2-m_2^2+s\right)}{32 \pi  s \left(m_{\hat{s}}^2-s\right)^2 \sqrt{\frac{\left(s-(m_1-m_2)^2\right) \left(s-(m_1+m_2)^2\right)}{s}}}.
\end{aligned}
\end{eqnarray}
The above equation is  merely a dominated term in this case, some terms with $\sin^{i}{\alpha}(i=2,3,4)$ are neglected.

\subsubsection{$\chi_2 \bar{\chi}_1\rightarrow \hat s\hat h$\rm (different scalar case)}
\begin{eqnarray}
\begin{aligned}
\sigma&=\frac{C_{\hat h\hat s\hat s}^2 y^2 \cos ^2\alpha \cos ^2 (2 \theta )  \left(-m_1^2+2 m_1 m_2-m_2^2+s\right) \sqrt{\frac{\left(s-(m_{\hat{h}}-m_{\hat{s}})^2\right) \left(s-(m_{\hat{h}}+m_{\hat{s}})^2\right)}{s}}}{16 \pi  s \left(m_{\hat{s}}^2-s\right)^2 \sqrt{\frac{\left(s-(m_1-m_2)^2\right) \left(s-(m_1+m_2)^2\right)}{s}}}.
\end{aligned}
\end{eqnarray}

We used the FeynCalc package~\cite{Shtabovenko:2020gxv,Shtabovenko:2016sxi,Mertig:1990an} to calculate relevant cross sections in this section.

\section{Decay width $\Gamma_0$}\label{appendix B}

\begin{figure}[t]
\begin{center}
\includegraphics[height=3cm,width=6cm]{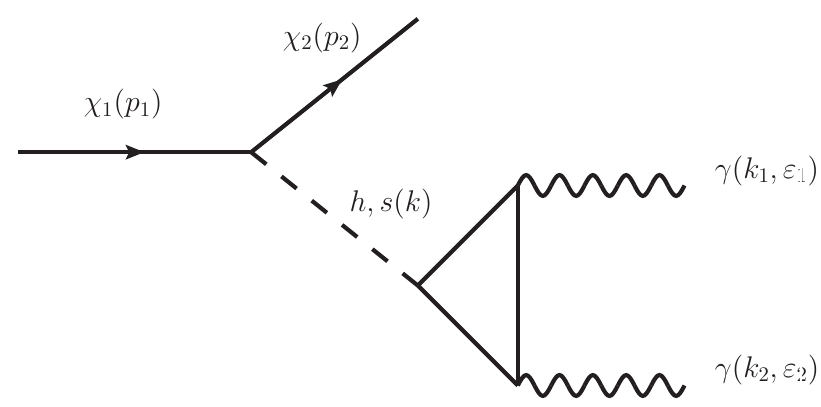}
\caption{Feynman diagram for the decay of $\chi_1\rightarrow\chi_2+2\gamma$.} \label{fig3}
\end{center}
\end{figure}

In this part, we discuss the decay of the  heavier $\chi_1$. The main contribution process of Feynman diagram is shown in Fig.\ref{fig3}.
%The one loop diagrams are the leading order contributions. The virtual top quark loop is the dominate contribution because of the large Yukawa coupling.
Through virtual top quark in the loop, the following effective operator $h\gamma\gamma$ can be given by\cite{Shifman:1979eb}

\begin{eqnarray}
\mathcal{L}_{h\gamma\gamma}=\frac{\alpha_{EM}}{2 \pi}Q^2_t\frac{g}{2m_w}(c^{'} \hat{h}-s^{'} \hat{s})F_{\mu\nu}F^{\mu\nu},
\end{eqnarray}
where $Q_t=2/3$ and $C_{eff}=\frac{\alpha_{EM}}{2\pi}Q^2_t \frac{g}{2 m_w}$ is effective coefficient. The $\alpha_{EM}=1/137$ is fine-structure constant. In addition, the couplings ${\chi}_1$, ${\chi}_2$ and $\hat{h}(\hat{s})$ are given in Eq.(\ref{coupling}).

The decay width of process $\chi_1\rightarrow \chi_2+2\gamma$ is calculated as follows. The amplitude of $\chi_1$ decay reads:
\begin{eqnarray}
\begin{aligned}
i\mathcal{M}=&\bar{u}(p_2){y\cos(2\theta)s^{'}\over\sqrt{2}}\gamma^{5}u(p_1){iC_{eff}c^{'}\over k^2-m^2_{\hat{h}}}2\left[(\varepsilon_1\cdot \varepsilon_2)(k_1 \cdot k_2)-(\varepsilon_1\cdot k_2)(\varepsilon_2 \cdot k_1)\right] \\
&-\bar{u}(p_2){y\cos(2\theta)c^{'}\over\sqrt{2}}\gamma^{5}u(p_1){iC_{eff}s^{'}\over k^2-m^2_{\hat{s}}}2\left[(\varepsilon_1\cdot \varepsilon_2)(k_1 \cdot k_2)-(\varepsilon_1\cdot k_2)(\varepsilon_2 \cdot k_1)\right],
\end{aligned}
\end{eqnarray}
where $u(p_1)$ and $u(p_2)$ denote the $\chi_1$ and $\chi_2$ fermion states, respectively. Next, we obtain the squared amplitude:
\begin{eqnarray}
\begin{aligned}
{1\over 2}\sum|\mathcal{M}|^2&=8\left({y\cos(2\theta)C_{eff}\over\sqrt{2}}\right)^2\left(\frac{s^{'2}c^{'2}}{(k^2-m^2_{\hat{s}})^2}+\frac{s^{'2}c^{'2}}{(k^2-m^2_{\hat{h}})^2}-{2s^{'2}c^{'2}\over (k^2-m^2_{\hat{s}})(k^2-m^2_{\hat{h}})}\right)\\
& \times \left[(p_1 \cdot p_2)-m_1 m_2\right](k_1 \cdot k_2)^2.\\ \label{mm}
\end{aligned}
\end{eqnarray}
We have,
\begin{eqnarray}
\begin{aligned}
\Gamma_0&=\frac{(2 \pi)^4}{2E_{p_1}}\int \delta^4(p_1-p_2-k_1-k_2)\frac{d^3\vec{p_2}}{(2 \pi)^3 2E_{p_2}}
\frac{d^3\vec{k_1}}{(2 \pi)^3 2E_{k_1}}\frac{d^3\vec{k_2}}{(2 \pi)^3 2E_{k_2}}{1\over 2}\sum|\mathcal{M}|^2.\label{width}
\end{aligned}
\end{eqnarray}
%And because
%\begin{eqnarray}
%\begin{aligned}
%I_{\alpha \beta}&=\int\frac{d^3\vec{k_1}}{ E_{k_1}}\frac{d^3\vec{k_2}}{E_{k_2}}k_{1\alpha}k_{2\beta}\delta^4(p_1-p_2-k_1-k_2)\\
%&=\frac{\pi}{6}((p_1-p_2)^2 g_{\alpha \beta}+2(p_1-p_2)_{\alpha}(p_1-p_2)_{\beta})\\
%&=\frac{\pi}{6}(k^2 g_{\alpha \beta}+2k_{\alpha}k_{\beta}).\label{Ia}
%\end{aligned}
%\end{eqnarray}
Combined with the above Eqs.(\ref{mm}),(\ref{width}), the $\Gamma_0$ is calculated to
\begin{eqnarray}
\begin{aligned}
\Gamma_0=&\int{1\over 64 \pi^3 m_1} 8\left(\frac{s^{'2}c^{'2}}{(k^2-m^2_{\hat{s}})^2}+\frac{s^{'2}c^{'2}}{(k^2-m^2_{\hat{h}})^2}-{2s^{'2}c^{'2}\over (k^2-m^2_{\hat{s}})(k^2-m^2_{\hat{h}})}\right)\\
&\times \left({y\cos(2\theta)C_{eff}\over\sqrt{2}}\right)^2\left[ (p_1 \cdot p_2)-m_1m_2\right](k_1 \cdot k_2)^2dE_{p_2}dE_{k_1}.
\end{aligned}
\end{eqnarray}
The relevant 4-momentum of initial and final state particles are written as,
\begin{eqnarray}
\begin{aligned}
p_1&=&(m_1,0), \ \ \ p_2=(E_{p_2},\vec{p_2}),\\
k_1&=&(E_{k_1},\vec{k_1}),k_2=(E_{k_2},\vec{k_2}).\\
\end{aligned}
\end{eqnarray}
Then, we obtain the following relationship,
\begin{eqnarray}
\begin{aligned}
&(p_1 \cdot p_2)=m_1 E_{p_2},\\
&k^2=(p_1-p_2)^2=m^2_1+m^2_2-2 m_1 E_{p_2},\\
&k_1\cdot k_2={1\over 2}(p_1-p_2)^2.
\end{aligned}
\end{eqnarray}
We need to calculate $E_{k_1}$ integral first, then by doing the kinematics one can show for any particular $E_{p_2}$ that $E^{\rm max/min}_{k_1}={1\over 2}\left( \left(m_1-E_{p_2}\right) \pm \sqrt{E^2_{p_2}-m^2_2}\right)$. Next, the limits of integration for the final $E_{p_2}$ integral are: $m_2<E_{p_2}<{1\over 2}({m_1+{m^2_2\over m_1}})$.
Finally, the expression of the decay width $\Gamma_0$ is
\begin{eqnarray}
\begin{aligned}
\Gamma_0&=\frac{y^2\cos^2(2\theta)\sin^2(2\alpha)C^2_{eff}}{16\pi^3}\frac{m^7_1}{240}\bigg[
(1-x^2)(1-5x-14x^2-145x^3-94x^4-145x^5\\
&-14x^6-5x^7+x^8)-120x^3(1+x+3x^2+x^3+x^4)\log{(x)}\bigg(\frac{1}{m^4_{\hat{s}}}+\frac{1}{m^4_{\hat{h}}}-{2\over m^2_{\hat{h}} m^2_{\hat{s}}}\bigg)\bigg],
\end{aligned}
\end{eqnarray}
where $x=m_2/m_1$.
\end{appendix}

%%%%%%%%%%%%%%%%%%%%%%%%%%%%%%%%%%%%%%%%%%%%%%
%%%%%%%%%%%%%%%%%%%%%%%%%%%%%%%%%%%%%%%%%%%%%%
%%%%%%%%%%%%%%%%%%%%%%%%%%%%%%%%%%%%%
\bibliography{refs}

\end{document}